%% file: radioAGN_in_mergers_with_LoTSS.tex
\documentclass[twocolumn]{aastex631}

\def\nii{{{\rm [N}\,{\sc ii}]}}
\def\as{$^{\prime\prime}$}
\def\am{$^{\prime}$}

\def\d{$^{\circ}$}

\def\msun{{$M_{\odot}$}}

\received{August 1st, 2024}
\revised{January 20th, 2025}
\accepted{February 4th, 2025}

\shorttitle{Radio AGN in Merging Clusters}
\shortauthors{Rickel \& Moravec, et al.}

\graphicspath{{./}{}}

\usepackage{hyperref}
\usepackage{amsmath}	
\usepackage{amssymb}

\usepackage{booktabs} 
\usepackage{graphicx}
\usepackage{multirow}
\usepackage{subfigure}
\usepackage{tabularx}

\begin{document}

\title{The Merging Galaxy Cluster Environment Affects the Morphology of Radio Active Galactic Nuclei}

\correspondingauthor{Emily Moravec} 
\email{emoravec@nrao.edu}
\input{authors.tex}

\begin{abstract}
\noindent
It has previously been found that the galaxy cluster environment can affect the fueling and evolution of active galactic nuclei (AGN). This work examines the effect of the merging cluster environment on the properties of radio AGN by comparing the radio morphology of cluster members in a sample of four merging and eight relaxed galaxy clusters at low redshift ($z$ $<$ 0.2). Using 144 MHz data from the LOFAR Two-meter Sky Survey and Zooniverse, we classify the radio morphology of the radio-detected cluster members using the following morphology classes: compact, compact extended, extended, jetted, and disturbed. We find that the merging cluster environment has a statistically significant, higher population proportion of disturbed (bent and head-tail) sources, indicating that the merging environment can affect the morphology of cluster radio AGN. We also investigate the number of AGN that are detected in the radio data only and the number that are detected in both the radio and optical data in mergers and nonmergers. We find that the merging cluster environment has a higher population proportion of AGN that are identified only as radio AGN compared to AGN that are identified as both radio and optical AGN. Overall, we find that the merging environment affects certain radio AGN (disturbed and only radio-identified AGN), but not all.

\end{abstract}


%%%%%%%%%%%%%%%%%%%%%%%%%%%%%%%%%%%%%%%%%%%%%%%%%%%%%%%%%%%%%%%%%%%%%%%%%%%%%%%%
\section{Introduction} \label{sec:intro}
It is now well established that dense environments such as galaxy clusters affect the properties and evolution of the galaxies within them (see references detailed below). There are a variety of interactions between the galaxies, the intracluster medium (ICM; a diffuse, hot, ionized plasma that is gravitationally bound to the cluster), and the cluster potential that are possible and prevalent in clusters that affect properties and evolution of galaxies. Some examples of the possible interactions are galaxy$-$galaxy harassment (repeated close encounters with neighbors; \citealt{Moore1996,Moore1999}), galaxy cluster potential tidal effects \citep{Fujita1998,Natarajan1998}, and galaxy interactions with the ICM, such as ram pressure stripping \citep{Gunn1972} and strangulation \citep{Larson1980}. These interactions can affect the star formation rates (SFR; \cite{Kauffmann2004,Fassbender2014,Barsanti2018}), active galactic nuclei (AGN) activity \citep{Galametz2009,Bufanda2017,Poggianti2017,Marshall2018, Mo2018}, morphology \citep{Dressler1980,Kauffmann2004,Sazonova2020}, and gas supply of the cluster galaxies \citep{Stroe2015,Poggianti2017,Jaffe2018,Cairns2019}.

One classic example of the effect of the cluster environment on its galaxies is the morphology$–$density relation, in which, with increasing density, the fraction of elliptical galaxies increases and the fraction of spiral galaxies decreases such that clusters typically have an overabundance of elliptical galaxies compared to the field \citep{Dressler1980,Kodama2001,Kauffmann2004,Sazonova2020}. Coupled to the morphology-density relation is the SFR, where more star-formation is seen in galaxies at low densities and less at high densities \citep{Kauffmann2004,Fassbender2014}. There is an evolution of these properties as a function of redshift, in that, with decreasing redshift, the fraction of ``red and dead" galaxies increases \citep{DeLucia2004,Rudnick2009} and the fraction of star forming galaxies decreases \citep{Margoniner2001,Poggianti2006,Saintonge2008}. Another example of the environment affecting cluster galaxies is the case of ``jellyfish" galaxies, which are galaxies typically located near the cores of the clusters that are experiencing ram pressure stripping, leaving a tail of gas behind them opposite to the direction of motion \citep{Fumagalli2014,Ebeling2014,Jaffe2018}.

The cluster environment is also known to affect the properties and evolution of cluster galaxies that host AGN. AGN are powerful astrophysical sources that are fueled by accretion of matter onto the supermassive black hole at the center of the galaxy and emit characteristic radiation signatures across the full electromagnetic spectrum \citep{Padovani2017}. AGN can be identified in one or multiple wavelengths such as X-ray, optical, infrared, or radio. At all redshifts \cite{Galametz2009} and \cite{Mo2018} find that there is an excess of radio-identified AGN in clusters, specifically at the center of clusters compared to the field, which is more pronounced at high redshift \citep{Galametz2009,Mo2020}. Similarly, pertaining to X-ray, optical, and infrared identified AGN, the number density of AGN continually increases as a function of redshift in that there is a deficit of these types of AGN near the cluster center at $z\lesssim0.5$ \citep{Pimbblet2013,Ehlert2014,Gordon2018}, but at $z\gtrsim0.5$ there is an excess of these types of AGN in clusters \citep{Ruderman2005,Galametz2009,Fassbender2012,Martini2013,Alberts2016,Bufanda2017} which is most pronounced near the cluster center \citep{Ruderman2005,Galametz2009,Fassbender2012,Alberts2016}. The cluster environment seems to affect many different properties of radio AGN in particular, such as the size, luminosity, and morphology of the radio AGN \citep{Prestage1988,Gendre2013,Ineson2015,Croston2019,Garon2019,Moravec2020a,Macconi2020,Shen2020,Morris2022}. Further, bent-tail radio galaxies are a very well-known type of radio AGN that showcase the effect of the environment on radio AGN produced by the ram pressure of the dense environment \citep{Miley1972,Begelman1979,ODonoghue1993,Hardcastle2005,Morsony2013}.

Galaxy clusters undergo cluster$-$cluster mergers (henceforth mergers\footnote{In this work we refer to cluster$-$cluster mergers as ``mergers," which we differentiate from the typical galaxy$-$galaxy mergers that many in the literature refer to as ``mergers."}) as a result of hierarchical structure formation in the Universe \citep{Voit2005,Springel2005,Boylan-Kolchin2009,Klypin2011,Pillepich2018,Planck2018VI}. Mergers have an immense and obvious impact on the properties of the ICM. For example, mergers cause the ICM distribution to be asymmetric as seen by X-ray \citep{Buote2002} and Sunyaev$–$Zel'dovich observations \citep{Mroczkowski2019}. Further, mergers drive intergalactic shocks and induce turbulence in the ICM \citep{Sarazin2002,Markevitch2006,Basu2016}. Lastly, mergers can create cold fronts in the ICM \citep{Sarazin2002,Owers2009,Roediger2013} and disrupt the cool cores of clusters \citep{Sarazin2002}.

Although the effect of mergers on the ICM is clear, the impact on the cluster galaxies is less so. There is contradictory evidence that mergers can either increase \citep{Owen1999,Miller2003,Hwang2009} or decrease \citep{Hwang2009, Mansheim2017} star formation in cluster galaxies. One study found that a dynamically younger cluster had a higher fraction of galaxies with a high SFR ($>$0.17 $M_{\odot}$ yr$^{-1}$, \citealt{Stroe2015}). There is also evidence that a brief period of enhanced, triggered star formation may be followed by subsequent quenching \citep{Stroe2015,Cairns2019}. These results are attributed to a disturbed ICM affecting the molecular gas reservoir of the cluster galaxies, with the disturbance exciting star formation and leading to a rapid consumption of the molecular gas \citep{Stroe2015,Cairns2019}.  

Recent work has found that the AGN population is impacted by cluster$-$cluster mergers as well. The radio AGN population in particular has been studied in depth in individual merging systems and found to have a higher frequency compared to the radio AGN frequency in more relaxed systems \citep[e.g.,][]{Miller2003,Moravec2020b}. Similarly, \cite{Noordeh2020} found enhanced X-ray AGN activity in the most dynamically disturbed of the seven clusters they studied. \cite{Bilton2020} find that mergers hold relatively kinematically younger AGN subpopulations that have recently coalesced into a common potential (similar to a first infall population of galaxies) compared to relaxed clusters. In general, \cite{Miller2003,Hwang2009} and \cite{Sobral2015} find that mergers can provoke AGN activity. However, the physical mechanism by which mergers instigate AGN activity and its effects on radio AGN properties are unknown.

In this paper, we explore the effect of mergers on the morphology of radio AGN. In \S\ref{sec:data}, we explain the sample and data used in this work. In \S\ref{sec:methods}, we describe the methods used to create images of the radio AGN and classify them. In \S\ref{sec:results}, we present the results of our investigations into the morphology and number of radio AGN in mergers versus nonmergers and compare them to optically identified AGN. In \S\ref{sec:disc}, we discuss the implications of our results. In \S\ref{sec:summary} we summarize our findings. Throughout this paper we use a cosmology in which $H_0=$ 70 kms$^{-1}$ Mpc$^{-1}$, $\Omega_{m}=$ 0.3, and $\Omega_{\Lambda}=$ 0.7.

\section{Sample and Data} \label{sec:data}

\subsection{The SDSS Galaxy Cluster Sample} \label{sec:data:Bilton}

The parent sample of galaxy clusters used for this study was obtained from \cite{Bilton2020}. 
This galaxy cluster sample was assembled with Sloan Digital Sky Survey Data Release 8 galaxies (SDSS DR8; \citealt{York2000,Aihara2011}), which were constrained by parameters found in the literature, and compiled into the X-Ray Cluster Database (BAX; \citealt{Sadat2004}).
The SDSS DR8 spectroscopic data are magnitude limited in the $r$band to $m_{{r}}\lesssim17.77$ \citep{Strauss2002}.
Furthermore, the MPA-JHU value-added catalog was cross-matched to the DR8 membership, in particular for their stellar mass estimates, in order to maintain completeness \citep{Kauffmann2003,Brinchmann2004,Salim2007}.

\begin{deluxetable*}{ccccccc}
	\centering
    \tablecaption{Galaxy Clusters Studied in This Work} \label{tb:cluster_info}
	\tablehead{\colhead{Cluster} & \colhead{R.A.} & \colhead{Dec.} &\colhead{$z$}& \colhead{$\mathrm{N}$} & 
    \colhead{RMS} & \colhead{$D_{\mathrm{state}}$} \\
    \colhead{} & \colhead{(J2000)} & \colhead{(J2000)} & \colhead{} & \colhead{} & \colhead{(mJy beam$^{-1}$)} & \colhead{}} 
	\startdata
   A1367 & 11:44:29.5   & $+$19:50:20.6   & 0.022  & 58  &0.245 & NM  \\
    A1656 & 12:59:48.7   & $+$27:58:50.5  & 0.023  & 96   & 0.117 & NM \\
    A1795 & 13:49:00.5  &  $+$26:35:06.8 & 0.062  & 46  & 0.127 & NM \\
    A2029 & 15:10:56.0  &  $+$05:44:41.0 & 0.077  & 98  & 0.307& NM \\
    A2061 & 15:21:15.3  &  $+$30:39:16.7 & 0.078  & 95 &0.0913& NM \\
    A2065 & 15:22:42.6  &  $+$27:43:21.5 & 0.073  & 87  & 0.123& NM \\
    A2069 & 15:23:57.9  &  $+$29:53:25.8   & 0.116  & 57 & 0.0765& NM \\
    A2199 & 16:28:38.5  &  $+$39:33:06.0 & 0.030  & 120  & 0.16& NM \\
    \midrule
    A2255 & 17:12:31.0 & $+$64:05:33.3 & 0.081  & 83     &0.18 & M\\
    A1991 & 14:54:30.2 & $+$18:37:51.1 & 0.044  & 63  & 0.116 & M\\
    A2033 & 15:11:28.1 & $+$06:21:43.7 & 0.082  & 41   & 0.184 & M\\
    A426 & 03:18:36.4  &  $+$41:30:54.2& 0.018 & 64   & 0.269& M\\
	\enddata
\tablecomments{Sample of clusters used in this analysis. Columns provide cluster name, coordinates, redshift ($z$), number of cluster members ($N$) identified by optical SDSS data in \cite{Bilton2020}, global RMS value of LoTSS image (mJy beam$^{-1}$), and dynamical state ($D_{\mathrm{state}}$, merging or nonmerging) for clusters from \cite{Bilton2020} covered by LoTSS. }
\label{tab:coord}
\end{deluxetable*}

\cite{Bilton2020} used the BAX galaxy cluster database to constrain the initial galaxy cluster sample to an X-ray luminosity range $1$ erg s$^{-1}$$ < L_{X} \leq 20 \times 10^{44}$ erg s$^{-1}$.
Applying these limits ensures that the most massive, well-assembled galaxy clusters are selected, while maintaining a sufficiently sized sample that can represent varying dynamical states within the finite limits of the $z$-space that can be feasibly observed.
After initial constraints were applied, the final cluster memberships for each cluster were then produced through surface caustics from the mass estimation methods of \cite{Diaferio1997} and \cite{Diaferio1999}.
These surface caustics vary as a function of projected radius $R$ from the cluster centre, thereby leading to the finalized membership being found within the confines of these caustics (see \citealt{Gifford2013,Gifford2013a}).
Once the cluster galaxy membership had been ascertained, \cite{Bilton2020} proceeded to distinguish between relaxed and unrelaxed states, or `nonmerging' and `merging,' respectively.
This was achieved via the incorporation of the \cite{Dressler1988} statistical test for substructure (also known as the $\Delta$-test), where the presence of substructure is used as a proxy for a galaxy cluster in a `merging' or unrelaxed dynamical state. 
The details of this statistical test and its calculation are elaborated on in \cite{Bilton2020}.

Utilizing the DR8 spectral lines for H$\alpha$ and [N$II$] $\lambda6584$, so-called `WHAN' diagrams can be used as a diagnostic to determine whether a cluster galaxy hosts an AGN or not \citep{CidFernandes2010,CidFernandes2011}. 
Specifically, the equivalent width of H$\alpha$, EW$_{\mathrm{H}\alpha}$, is compared to the logarithmic ratio of log$_{10}$(\nii/H$\alpha$).
The reduction of the number of emission lines for a WHAN diagram to two allows for mitigation against the signal-to-noise (S/N) $\geq3$ requirements for all emission lines, especially when compared to the Baldwin$-$Philips$-$Terlavich (BPT) diagnostic plot's use of four emission lines \citep{Baldwin1981}. We refer to these WHAN-selected AGN as optical AGN in the rest of this work. Furthermore, to reduce the number of interloping star-forming galaxies and low ionisation emission regions, galaxies are classified as AGN if they possess emission ratios and strengths of log$_{10}$(\nii/H$\alpha$)$\geq-0.32$ and EW$_{\mathrm{H}\alpha}$ $\geq$ 6\AA \ respectively.
%[N\textsc{\footnotesize~II}]

During our initial analysis of the cluster member catalogs from \cite{Bilton2020}, we discovered duplicate sources. We determined that these duplicates were a product of the fact that the SDSS DR8 spectroscopic catalog contains multiple measurements for some sources (along the plate overlap regions). We removed the 143 duplicate sources, which leaves us with a total of 2298 objects. 

In addition to the duplicates found within each cluster, we also found 42 sources that were counted as members in both the A2029 and A2033 member catalogs. We note that \cite{Walker2012} show X-ray evidence of overlap between these two systems. To remedy this, for each of these 42 duplicate members we calculate a statistic, $C$, which is based on both projected separation and relative velocity \citep{Smith2004}. The statistic $C$ is defined as 
\begin{equation}
    C = (cz - cz_{\mathrm{cl}})^2/\sigma_{\mathrm{cl}}^2 - 4 \log (1 - R/R_{\mathrm{cl}})
\end{equation}
where $c$ is the speed of light, $z$ is the redshift of the galaxy, $z_{\mathrm{cl}}$ is the cluster redshift, $\sigma_{\mathrm{cl}}$ is the velocity dispersion of the cluster, $R$ is the projected radius of the galaxy from the cluster center in r$_{200}$, and $R_{\mathrm{cl}}$ is the radius of the cluster in units of r$_{200}$. 
We then assign the member to the cluster that minimizes $C$ \citep{Smith2004}. Minimizing $C$, 37 of the duplicates are assigned to A2029 and 5 are assigned to A2033.

\subsection{LoTSS Data} \label{sec:data:lotss}
To examine the radio characteristics of the cluster members, we used data from the LOFAR Two-meter Sky Survey (LoTSS: \citealt{Shimwell+2017}). LoTSS provides a resolution of 6\as\ with a central frequency of 144 MHz and a median RMS sensitivity in public data of 83 $\mu$Jy beam$^{-1}$. We made use of the current public data release of LoTSS which is Data Release 2 \citep{Shimwell2022}; in addition we made use of data processed since that release, which will be made publicly available in LoTSS Data Release 3.
 
Out of the 33 clusters from \cite{Bilton2020}, 13 were covered by the LoTSS survey, 9 nonmerging and 4 merging. Of these 13 clusters, one, A119, was only half covered by the SDSS and was therefore discarded from the analysis. As a result, our final sample consisted of 12 clusters fully covered by LoTSS and SDSS (see Table \ref{tab:coord}). For each of these clusters we made a mosaicked image that combined all of the available LoTSS pointings covering that cluster, maximizing the sensitivity of the images used for this study.

\subsection{Completeness and Redshift Distribution} \label{sec:data:z}
It is known that SDSS completeness drops to $\approx$65\% in denser fields such as a galaxy cluster \citep{Yoon2008}, due to there being more targets than available fibers. To investigate the completeness of our sample, we compare the average mass of the galaxy clusters and the number of cluster members across both dynamical states. When examining the \cite{Bilton2020} SDSS catalogs of the LoTSS covered clusters, we find an average of 62 members per merging cluster and 82  per nonmerging cluster. When we match these averages with average cluster halo mass, we find that mergers have 5.4 x$10^{14}$\msun\ and nonmergers have 6.3 x$10^{14}$\msun. Cluster mass, $M_{200}$, is determined in \cite{Bilton2020} from the density profile  $\rho(r) = 3M(r)/4\pi r^{3}$, where  $\rho(r)$ is specifically 200 times the critical density of the flat Universe.  If we compare the number of cluster members and halo masses of the clusters, we find that merging clusters relative to nonmergers have about 75\% of the members and 80\% of the mass. The similarity of these values is suggestive that there is not a significant selection bias.

In addition to survey completeness, we investigate the redshift distribution within each merging and nonmerging populations to ensure that they are well matched. Given this work is rooted in morphology, there exists a potential bias of a morphology class being mistaken for another owing to resolution issues coming from redshift differences. For the nonmergers, we find the mean $z$, median $z$, and interquartile range to be 0.06, 0.07, and 0.05, respectively. For mergers, we find the mean $z$, median $z$, and interquartile range to be 0.06, 0.06, and 0.05, respectively. The redshifts are well matched between the cluster dynamical state populations, and we do not expect redshift to bias morphological classifications within each population.

\section{Methods} \label{sec:methods}
In order to determine whether or not the merging cluster environment affects the radio morphology of AGN and cluster galaxies, the radio sources need to be classified according to their morphology. To do this, we created images (\S\ref{sec:cutouts}), devised a classification system (\S\ref{sec:classifications}), visually classified the radio sources using Zooniverse (\S\ref{sec:zooniverse}), aggregated the votes from visual inspection and defined consensus levels (\S\ref{sec:consensus_level}), determined which sources were radio AGNs (\S\ref{sec:agnVsSF}), and performed statistical analyses (\S\ref{sec:meth:beta}). 

\subsection{Cutouts} \label{sec:cutouts}
In preparation for classification, we needed to create cutouts of each source with radio emission. To do this, we first obtained LoTSS images of each cluster that were 2 $r_{200}$ $\times$ 2 $r_{200}$. Once the images were obtained, we made ``cutouts" or images that were zoomed in on each cluster member. The dimensions of these cutouts varied in size from 12\as\ $\times$ 12\as\ for compact emission to 60\as\ $\times$ 60\as\ in order to be able to showcase the extended emission of the source. 

Since there were a large number of sources and respective cutouts to be made, there was a necessity to be able to create bulk cutouts for the classification process. Thus, we calculated and used a ``global" cluster noise (henceforth referred to as RMS or $\sigma$) for generating contours for the radio sources. To determine the global RMS for each LoTSS image, we made use of SAOImageDS9 (\citealt{ds9}). For each mosaicked cluster image, we selected an emission-free region and drew a circle $\approx$4{\am} in diameter within this region. From this circle, we extracted the RMS in mJy beam$^{-1}$ from the statistics calculated automatically in DS9. For each cluster, we used this global cluster RMS to produce the contours for the cluster member cutouts (reported in Table \ref{tab:coord}). 

The method of calculating the RMS and contour levels displayed for the cutouts can influence the appearance and thus the morphology classification of the source. Given our dataset, which comprises over 1000 sources, we were confronted with a trade-off between efficiency and sensitivity. We investigate whether the morphology classification of a source is RMS (and corresponding contour levels) dependent by selecting a handful of sources (14) showing faint emission potentially related to the source but not captured with contours and testing the effect of using different RMS values. For these sources, we created new cutouts using varying RMS and contour levels. We compared the original, global RMS and contour level for each image to 1) the local RMS (determined from an $r=100$\as\ circle of an emission-free region in the cutout) with the original 4$\sigma$ $\times$ 2$^n$ base for contours; 2) the original global (cluster) RMS but with a 3$\sigma$, 4$\sigma$, 5$\sigma$, 8$\sigma$ contour levels; and 3) the local RMS with 3$\sigma$, 4$\sigma$, 5$\sigma$, 8$\sigma$ contour levels. We find no significant deviation from the original classification results when comparing the new RMS and contour levels. As a result, we are confident that our choice of using a global RMS is robust.

\begin{figure*}
\centering
  \includegraphics[width=18cm,height=22cm]{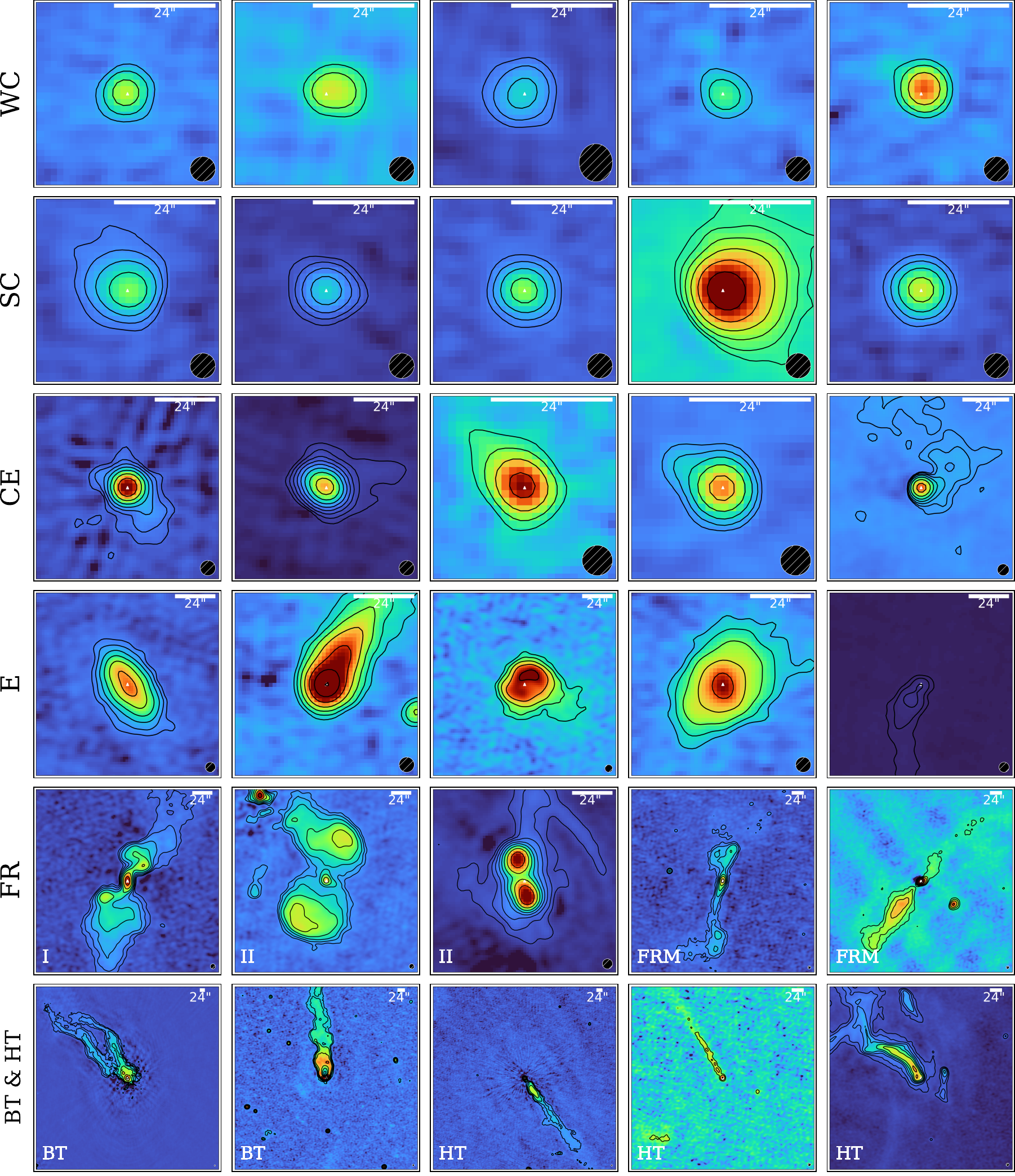}
    \caption{Examples of all classifications. See \S\ref{sec:classifications} for classification definitions. Representative images use the local RMS (determined from an $r = 100$\as\ circle of an emission-free region in the cutout) with the original 4$\sigma$ $\times$ 2$^n$ base for contour level. These images may differ slightly from those shown to classifiers. For all panels, the LoTSS beam is displayed in the lower right corner.}
  \label{fig:morph_classes}
\end{figure*}

To aid with the classification process, we displayed contours in the cutouts that were log-based starting with a 4$\sigma$ base determined by the cluster RMS, as described above. Thus, the contours begin with 4$\sigma$ and increase by factors of 2$^n$ where $n=$ 1, 2, 3, etc. To optimize visualization for many hundreds of images, we employed a square root scaling technique in \texttt{aplpy} and scaled the images as a percentage of the maximum pixel value (\texttt{pmax} = 100). To see all of these parameters in use, see Figure \ref{fig:morph_classes}.

\subsection{Radio Morphology Classifications} \label{sec:classifications}
After the cutouts were made, one team member visually categorized the sources as having a radio detection or nondetection. A source was defined as having a radio detection if it had $\geq$8$\sigma$ detection at the cluster member coordinate from the \cite{Bilton2020} catalog, which manifested visually as two contours (at a 4$\sigma$ base). We chose 8$\sigma$ as the threshold for a detection for the following reasoning. Because we used a global, cluster RMS to create the contours for the cluster members, there are a few cases where 4$\sigma$ contour encapsulates the noise instead of the morphology of the source, making the radio morphology of the source difficult to classify. Since we were classifying the sources as detection or nondetection visually (to be consistent with the visual radio morphology classification), we increased the base of detection to the next displayed contour, which in this case is 8$\sigma$ (because we define the contours as 4$\sigma$ $\times$ 2$^n$ where $n=$ 0, 1, 2, etc.) in order to be confident that the contours were displaying radio emission and not noise.

After visual inspection of the cutouts of the \cite{Bilton2020} sample of 2298 galaxies (1670 from nonmergers, 628 from mergers), we find that 191 had a radio detection. These radio detection images were uploaded to Zooniverse\footnote{\href{https://www.zooniverse.org/}{Zooniverse website}} for visual morphological classification by five team members (M.R., E.M., Y.A.G., L.E.B., J.C.S.P.). 

Creating a classification scheme was an iterative process. There was an initial, baseline classification scheme that was created during preliminary work based on the spatial distribution, extension, and strength of the radio emission of the source. After testing this initial classification scheme, we converged on a final morphology scheme that we used on Zooniverse and follows (in order of increasing level of disturbance):
\begin{itemize}
    \item \textbf{\textit{Weak Compact (WC)}}: Emission (a) is roughly circular, (b) is on a scale of less than or roughly equal to 12{\as} (2 times the resolution of LoTSS), and (c) has two or three contours (8$\sigma$-16$\sigma$ detection).
    \item \textbf{\textit{Strong Compact (SC)}}: Emission that (a) is roughly circular, (b) is on a scale of less than or roughly equal to 12{\as} (2 times the resolution of LoTSS), and (c) has four or more contours ( $\geq $ 32$\sigma$ detection). There may be sources that have 4 contours and are roughly circular but are larger than 12{\as} (up to 30{\as} in size); these are still strong compact.\footnote{This condition is meant to catch strong compact sources that only differ from the original classification by being larger than 2 times the LoTSS resolution.}
    \item \textbf{\textit{Compact Extended (CE)}}: Emission has a compact (roughly circular and with a size of less than or roughly equal to 30{\as}) core with a singular source of extended emission that is on the order of or greater than 12{\as} (2 times the resolution of LoTSS).
    \item \textbf{\textit{Fanaroff and Riley Type I (FRI)}}: A source that follows the canonical FR I type morphology \citep{Fanaroff1974} and is not bent.
    \item \textbf{\textit{Fanaroff and Riley Type II (FRII)}}: A source that follows the canonical FR II type morphology \citep{Fanaroff1974} and is not bent.
    \item \textbf{\textit{Fanaroff and Riley Type Morphology (FRM)}}: A source with clear two-sided non-bent jets, but FR classification is unclear.
    \item \textbf{\textit{Extended (E)}}: Emission that (a) is clearly on a much larger scale than that of the compact sources on the order of or much greater than 5 times the resolution of LoTSS ($~$30") but is not an FR I, FR II, bent-tail, or head-tail source and (b) is irregular in shape (not circular).
    \item \textbf{\textit{Bent Tail (BT)}}: A source in which the jets are bent to some degree. Either or both jets could be slightly or extremely bent.
    \item \textbf{\textit{Head Tail (HT)}}: An extreme bent-tail source where the two jets/tails are indistinguishable.
    \item \textbf{\textit{Undetermined}}\footnote{In the Zooniverse project this classification was called ``Uncertain," but the name of the classification was changed to Undetermined for this paper to avoid confusion with other types of uncertainty. The change in classification class name does not change any results.}: There is obviously a source in the anticipated location, but the classification is uncertain.
    \item \textbf{\textit{Bad Data}}: When there are clearly imaging artifacts interfering with the source.
    \item \textbf{\textit{No Detection}}: Those that have no emission near the position of the cluster member (indicated by a triangle).
\end{itemize}
Sources are classified as ``Undetermined" if their morphology does not follow the definition of any provided classifications. The classification system describes the majority of our sample. We note that no morphological classification scheme will be able to exhaustively describe all of the possible complex morphologies. Exemplary sources of each classification are shown in Figure \ref{fig:morph_classes}.

\subsection{Zooniverse Project}\label{sec:zooniverse}
To obtain robust classifications, five team members visually classified each source. For image sharing and collaborative classification, we made use of Zooniverse, a widely recognized online tool designed for projects reliant on classification-based tasks \citep{Lintott2008}. We established a private project and invited collaborators who are experts in fields relating to the scientific scope of this paper to participate. Upon joining the project, each classifier was presented with a randomly selected radio image from the dataset and prompted to classify it based on the classification criteria described in \ref{sec:classifications}. The classifiers were blind to the dynamical state of the parent cluster (merging vs. nonmerging) and to which cluster each source came from. The accumulated results from these five evaluators were incorporated into this study for analysis. We note that each classifier was allowed to give each source at most one radio classification and multiple classifications were not allowed.

\subsection{Consensus Level}\label{sec:consensus_level}
We aggregate the votes from Zooniverse. Following the aggregation of votes from Zooniverse, we define multiple levels of consensus. Consensus level was determined by the number of classifiers (of the five total) that agreed on a source classification. For example, in order to create a group of sources for which we were extremely confident in their classification, we define “consensus level 4” as four or five out of five classifiers agreeing on a classification. Next, to retain a majority classification but include as many sources as possible, we define “consensus level 3” as three of the five classifiers agreeing on a classification. Consensus level 3 also includes classifications that had a consensus level of 4. Lastly, we define a ``nonconsensus" class where 2 or fewer classifiers agreed on a classification. There are 71 consensus level 5 sources, 120 consensus level 4 sources, 166 consensus level 3 sources, and 25 nonconsensus sources.

We find that the results in this work remain consistent when using consensus level 3 or 4. A comparison between these two consensus levels is described in Appendix \ref{sec:app:Consensus3vs4}. Because the overall trends are the same in both samples, we use the consensus level of 3 to serve as the standard consensus level for the analysis in this work.

There were 9 sources out of the total of 166 consensus level 3 sources that were removed from the sample. There were six poor-image-quality sources that were removed. Any sources that appeared to have significant artifact contamination or very low strength as compared to the background image were classified as bad data (one) or no detection (five), respectively, and removed. Additionally, we omit the single source classified as undetermined in the consensus level 3 sample, as we want to investigate sources with a distinct radio morphology. We omit these sources (seven) from the analysis, for a total of 159 consensus level 3 sources. However, for the sake of completeness, we note here that after the radio AGN delineation process we removed two extended sources (the reason for which is discussed at length at the end of Section \ref{sec:agnVsSF}), for a total of 157 sources. We refer to these sources as the ``Zooniverse results," and provide a catalog of these 157 sources with this work for download (see Appendix \ref{sec:app:raw_zooniverse}).

\subsection{AGN vs. Star Forming in Compact Radio Morphologies}
\label{sec:agnVsSF}

\begin{figure}
    \centering
    \includegraphics[width=\columnwidth]{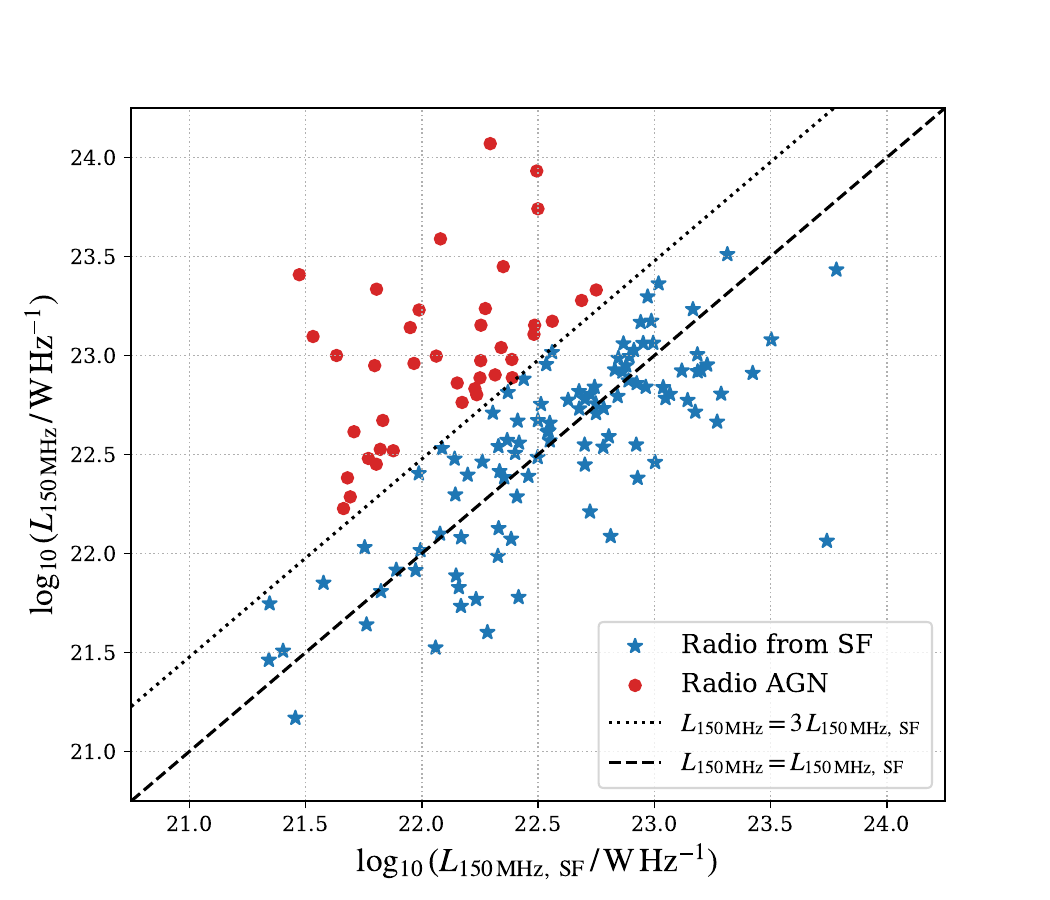}
    \caption{Expected radio emission due to star-formation, $L_{150\,\text{MHz},\ \text{SF}}$, versus $150\,$MHz for our compact radio sources.
    The black dashed line shows where the measured luminosity is equal to the luminosity expected owing to star-formation, while the dotted line shows \text{$L_{150\,\text{MHz}}=3\,L_{150\,\text{MHz,}\ \text{SF}}$}, which we use to identify sources where the radio emission is due to an AGN (red circles) as opposed to possibly being the result of star-formation (blue stars).}
    \label{fig:sf-v-agn}
\end{figure}

We further filter the \text{159} Zooniverse consensus 3 level sources from \S\ref{sec:consensus_level} to only those that are radio AGN in the following way. For radio galaxies with jetted and disturbed morphologies (11 total from the FR I, FR II, FRM, HT, and BT categories) it is clear that the radio emission is the result of an AGN. However, for compact radio sources where the AGN jet is unresolved, we must consider another possibility for the radio emission. In galaxies that do not host an AGN, star-formation produces radio emission, primarily as the result of synchrotron emission from supernova shocks and thermal emission from \textsc{H}\textsc{ii} regions \citep{Condon1992}. Generally, radio emission from star-formation is relatively low luminosity, dominating the local radio luminosity function at $L_{1.4\,\text{GHz}}<10^{23}\,\text{W}\,\text{Hz}^{-1}$ \citep{Best2005}. In order to analyze the impact of the cluster environment on radio AGN, it is therefore important to establish that the radio emission for compact objects is indeed the result of an AGN and not star-formation.

For the simple-morphology radio sources in our sample, namely those classified as WC, SC, or CE (141 sources), we estimate the expected $150\,$MHz emission due to star-formation based on their measured SFRs using the relation of \citet{Gurkan2018}. For these sources the $150\,$MHz flux density is measured from the LoTSS images using Mohan \& Raffergy’s \texttt{PyBDSF} \citep{Mohan2015}. Additionally, the galaxies in our sample are in the SDSS DR8 spectroscopic catalog and have SFR measurements from the MPA-JHU catalog \citep{Brinchmann2004}. By comparing the measured radio luminosity, $L_{150\,\text{MHz}}$, with the radio luminosity expected from star-formation, $L_{150\,\text{MHz,}\ \text{SF}}$, we can thus identify the likely source of the radio emission in our galaxies (see Figure \ref{fig:sf-v-agn}). The \citet{Gurkan2018} relation between SFR and radio luminosity has a typical uncertainty of 1.68. To account for this, and given that we are looking to only include radio AGN in our analysis, we adopt a conservative approach and consider sources with \text{$L_{150\,\text{MHz}} > 3\,L_{150\,\text{MHz,}\ \text{SF}}$} as radio AGN (red circles in Figure \ref{fig:sf-v-agn}). We find that 103 of our simple-morphology radio sources have radio emission that can be explained by star formation, leaving us with 38 radio AGN in our sample that have WC, SC, or CE  morphologies.

Further, of the seven E morphologies, three were easily delineated as radio AGN or star forming using the flux cutoff from their LoTSS catalog value. Two sources are visually determined to possess a degree of jetting on this second pass and therefore are classified as radio AGN. We honor the original classification of E for these sources, as poorly resolved jetted sources classified as E is a possible uncertainty discussed later in Section \ref{sec:results:combinedClassifications}. Of the remaining two E sources, one did not have a LoTSS catalog entry and could not be detected by Mohan \& Raffergy’s \texttt{PyBDSF}. Because this source does not have a reliable flux measurement and cannot be placed on the \citet{Gurkan2018} relation, we exclude this source. Through optical counterpart analysis, the last E source appears to be the overlap of two spiral galaxies but with only one LoTSS entry. Because it is not possible to determine which galaxy is contributing to the radio emission, we also drop this source from the analysis. Since these two sources were not able to be classified as a radio AGN or not, we entirely remove them from the sample, reducing the sample from 159 sources to 157 sources. 

Using this analysis to separate out the radio AGN, we find that 43 sources are radio AGN that have WC, SC, CE or E morphologies in consensus level 3. Thus, as a result, from the 157 Zooniverse sources, the final sample of sources that we use for analysis in the following work is 54 radio AGN consensus level 3 sources (11 FR I, FR II, FRM, HT, or BT + 43 WC, SC, CE or E).

While ram pressure stripping can boost the radio luminosity of a source, this is most extreme (approximately a factor of 3) at $L_{144} < 10^{21}$ W Hz$^{-1}$, with much lower levels of radio luminosity boosting at $L_{144} < 10^{22}$ W Hz$^{-1}$ (see Figure 9 of \citealt{Roberts2021}). All of our selected radio galaxies have $L_{144} > 10^{22}$ W Hz$^{-1}$ and thus are subject to low levels of radio luminosity boosting by ram pressure stripping. Furthermore, our radio galaxy sample is chosen to have radio luminosities of at least 3 times the expected radio luminosity owing to star formation (see Fig. \ref{fig:sf-v-agn}). As such, we are confident that all of our selected radio galaxies host a radio-loud AGN. 

\subsection{Statistical Methods}
\label{sec:meth:beta}
We want to compare the proportion of observed radio morphology classes in mergers and nonmergers within the uncertainties. To do this, we express our results as proportions ($P = k$/$n$), where $k$ is the number of radio AGN of a particular radio morphology in mergers or nonmergers and $n$ is the total number of radio sources for the respective dynamical state.

Given the small number statistics and asymmetric errors on the proportions measured, care is necessary in quoting and defining uncertainties. We take a Bayesian approach and give credible intervals on the true proportion $P$ given the observed $k$ and $n$. For a Jeffreys prior, the posterior probability distribution of $P$ is a Beta distribution,
\[
p(P|k,n) = \mathrm{Beta}(k + \frac{1}{2}, n - k + \frac{1}{2})
\]
and hence the credible interval can be taken to be the relevant percentile of the Beta distribution (e.g. 16\% and 84\% for a $1\sigma$ range as used here). It is conventional to set the lower range to 0 if $k=0$ and the upper range to 1 if $k=n$ (the Jeffreys interval, which is indicated as the lower and upper uncertainties in Figure \ref{fig:binom_combstat}), as implemented in the {\sc astropy.stats} function {\sc binom\_conf\_interval}.

It is important to note that, given the asymmetric nature of these errors and the fact that the posterior distribution about $k$/$n$ is far from being a Gaussian, it is not possible simply to combine the quoted `errors' (credible interval) in order to compare two different proportions with one another, or to multiply by a constant to get a $3\sigma$ confidence range. If we observe two proportions $P_1(k_1,n_1)$ and $P_2(k_2,n_2)$ where $k_1$/$n_1 < k_2$/$n_2$, then the Bayesian way of asking the question `Is $P_2$ significantly greater than $P_1$' is, `Is there a high posterior probability that $P_2 - P_1 > 0$?' This could be found exactly by convolving the two posterior distribution functions, but for simplicity we simply subtract samples drawn from $P_1$ and $P_2$, which allows us to estimate this posterior probability to adequate precision (listed as Pr(M$>$NM) in Table \ref{tab:combinedMorphologyStats}). We map the Pr(M$>$NM) to a statistical significance level using a one-tailed test by inverting the cumulative distribution function of a Gaussian. Although an approach framed as `Is there a difference?' or in other words testing whether Pr(M$>$NM) or Pr(M$<$NM), may seem less biased, there are cases where the proportion is 0, making it unreasonable to test for any direction other than greater. For consistency, we maintain a one-tailed approach throughout the analysis. These statistical significances are quoted and interpreted in Section \ref{sec:results:combinedClassifications}, taking the standard approach that 68\%, 95\%, and 99.7\% correspond to 1$\sigma$, 2$\sigma$, and 3$\sigma$, respectively.

\section{Results} \label{sec:results} 
After determining consensus and radio AGN delineation, we analyze the trends in our data based on morphology (see \S\ref{sec:results:combinedClassifications}) and AGN classification (see \S\ref{sec:results:optragn}).

\begin{table*}
    \caption {Radio AGN Morphology Results for Combined Morphology Classes.}
    \centering
    \renewcommand{\arraystretch}{1} % Adjust row spacing
    \begin{tabularx}{0.64\textwidth}{X*{8}{>{\centering\arraybackslash}X}}
        \toprule
         & C & CE & E & Jetted & Disturbed \\
        \toprule
        Tot$_{\mathrm{NM}}$ & 22 & 6 & 5 & 3 & 1\\
        Tot$_{\mathrm{M}}$ & 9 & 1 & 0 & 2 & 5 \\
        \midrule
        P$_{\mathrm{NM}}$ & $0.59^{+0.08}_{-0.08}$  & $0.16^{+0.07}_{-0.05}$  & $0.14^{+0.06}_{-0.05}$  & $0.08^{+0.05}_{-0.04}$ & $0.03^{+0.04}_{-0.02}$ \\  
        P$_{\mathrm{M}}$ & $0.53^{+0.12}_{-0.12}$  &$0.06^{+0.07}_{-0.04}$  & $0^{+0.04}$ & $0.12^{+0.09}_{-0.06}$  & $0.29^{+0.11}_{-0.10}$ \\
        \midrule
$\mathrm{Pr(M > NM)}$ & 0.325 & 0.146 & 0.039 & 0.68 & 0.998 \\         
        \bottomrule
    \end{tabularx}
    \label{tab:combinedMorphologyStats}
    \tablecomments{Column headers are shorthand for the combined classification groups: compact (WC+SC), compact extended, extended, jetted (FRI+FRII+FRM), and disturbed (HT+BT). Rows describe the total radio AGN and proportion of the classification within nonmergers and mergers using a beta-binomial Jeffreys Bayesian prior to determine the upper and lower bound uncertainty. $\mathrm{Pr(M > NM)}$ corresponds to the posterior probability that a specific classification is greater in mergers compared to nonmergers, discussed in Section \ref{sec:meth:beta}.} 
\end{table*}

\subsection{Combined Classifications and Results} \label{sec:results:combinedClassifications}
Though the individual classifications defined in \S\ref{sec:classifications} trace unique behavior across sources, we establish combined classification groups to trace general radio morphology classification trends between the cluster dynamical states. We define the following combined radio morphology classification groups based on morphological similarity:
\begin{itemize}
\item \textbf{\textit{Compact}}: This group of classifications combines the sources that are physically compact: weak compact (WC) and strong compact (SC) sources. The sources are grouped together because they fundamentally share the same characteristic in terms of scale of emission, only differing in emission strength.
\item \textbf{\textit{Compact extended}}: This group of classifications includes only compact extended (CE) sources. We keep CE separate from the compact objects, as they exhibit a single extension but differ from clearly extended, as they have a clear compact core.
\item \textbf{\textit{Extended}}: This group includes only the sources in the extended (E) category. In this sample, sources categorized as extended possess a great range in morphology. 
\item \textbf{Jetted}: This group combines sources that have FR I, FR II, or FRM classifications. We combine these classes, first in order to increase the number of sources in this class, and second because they both display unbent jets that have clear two-sided jet morphology. It it not important to distinguish between FR I and FR II for this work because they both have unbent, two-sided jets that are similar morphologically compared to the other types of morphologies considered in this work (e.g., compact, bent, etc.)
\item \textbf{\textit{Disturbed}}: This group of classifications combines bent-tail (BT) and head-tail (HT) sources. Tailed sources are indicative of ram pressure stripping as a galaxy navigates the dense ICM. The degree to which a source is bent is beyond the scope of this work; thus, we simply combine all sources that have jets that are bent to some degree to differentiate the bent from unbent radio morphologies.
\end{itemize}

\begin{figure}
    \centering
\includegraphics[width=\columnwidth]{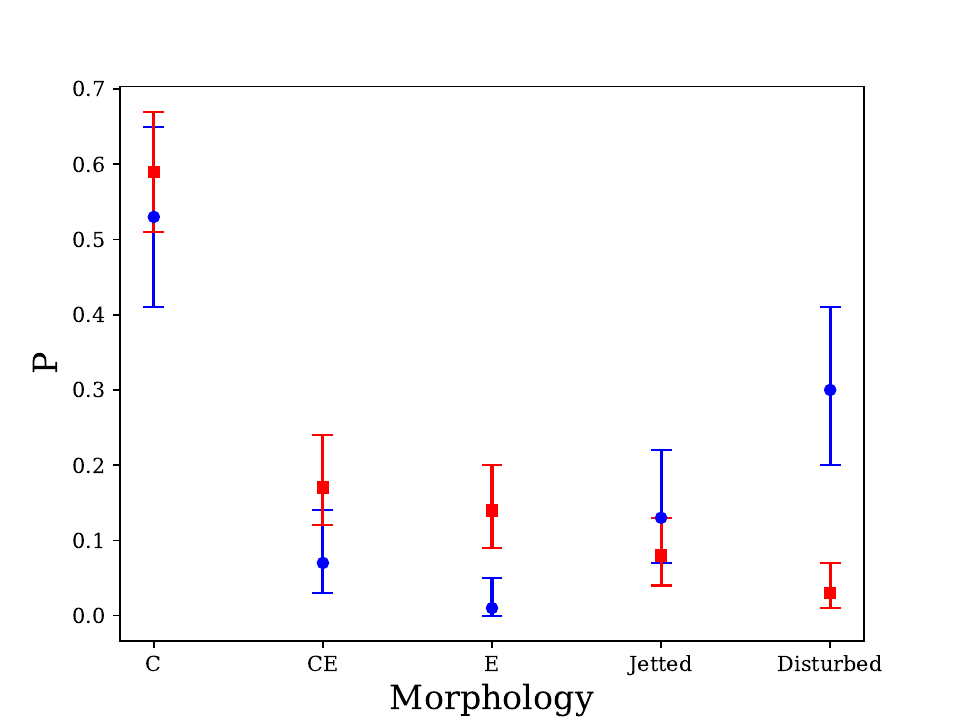}
    \caption{Binomial proportions (P) for all combined classifications visualized for nonmergers (red squares) and mergers (blue circles) referenced in Table \ref{tab:combinedMorphologyStats}. Upper and lower bound uncertainties are the 84th and 16th percentile value, respectively, from the beta-binomial distribution described in Section \ref{sec:meth:beta}.}
    \label{fig:binom_combstat}
\end{figure}

Enumerated in Table \ref{tab:combinedMorphologyStats} and visually in Figure \ref{fig:binom_combstat}, we report the proportion and uncertainties (described in \S\ref{sec:meth:beta}) of the combined morphology classifications. For the compact category of sources, in nonmergers, we observe population proportions of $0.59^{+0.08}_{-0.08}$ compared to $0.53^{+0.12}_{-0.12}$ in mergers. 
Similarly in the CE category, in nonmergers, the value is $0.16^{+0.07}_{-0.05}$  compared to $0.06^{+0.07}_{-0.04}$ in mergers. Using the posterior probabilities of the C and CE across mergers and nonmergers, the merging population does not exhibit statistically significantly greater proportions of either C or CE sources compared to nonmergers. In other words, for sources in the C and CE categories their corresponding Pr(M$>$NM) are 0.325 and 0.146, respectively, which both map to the equivalent of $<1\sigma$. The $\sigma$ calculation follows from mapping Pr(M$>$NM), where Pr(M$>$NM) is equivalent to (M $-$ NM $>$ 0) divided by N samples, to the corresponding $\sigma$ values discussed in Section \ref{sec:meth:beta}.

Interestingly, we find E sources only in nonmergers. In nonmergers, there is a population proportion of $0.14^{+0.06}_{-0.06}$ compared to $0^{+0.04}$ in mergers. Mergers do not display a significantly greater ($\sigma<1$) population proportion of E sources as compared to the nonmergers. Sources classified as E range greatly in morphology. This classification aimed to describe all extended, amorphous emissions that could not be unambiguously classified as compact or jetted. Based on a preliminary, quick optical counterpart examination by eye, E sources may arise from a variety of physical mechanisms, including but not limited to starburst face-on spirals, jellyfish galaxies, and cluster member mergers. The physical reasoning for heightened E sources in nonmergers is unclear, and a more detailed follow-up with optical counterparts would pose an interesting future direction. The origin of the emission mechanism for E sources is beyond the scope of this paper and requires a multiwavelength approach for robust analysis. 

For the jetted category of sources (the jetted sources encompass all clear FR-type morphologies), we find the observed population proportion for mergers is $0.12^{+0.09}_{-0.06}$ and that for nonmergers to be $0.08^{+0.05}_{-0.04}$. And using the posterior probabilities of the jetted sources across mergers and nonmergers, we find that the merging population does not exhibit a statistically significant greater proportion of jetted sources (Pr(M$>$NM) of 0.68 which is the equivalent of $1\sigma$).

Lastly, we find that the merging systems have a greater proportion of disturbed sources than nonmergers, within the error. We find that mergers and nonmergers have a population proportion of $0.29^{+0.11}_{-0.10}$ and $0.03^{+0.04}_{-0.02}$, respectively. And using the posterior probabilities of the disturbed sources across mergers and nonmergers, we find that the merging population does exhibit a statistically significant greater proportion of disturbed sources as compared to nonmergers (Pr(M$>$NM) of 0.997 which is equivalent to 3$\sigma$). After visually re-examining the images of the disturbed sources, we find that they appear to be both bent and head-tails, which are jetted sources that have been affected by ram pressure ($P_{\mathrm{ram}} = \rho v^2$), causing the jets to become bent to some degree \citep{Miley1972,Owen1976,Begelman1979,ODea1985}.

In summary, we find that mergers do not exhibit greater proportions of C sources, CE sources, and jetted sources while they contain no E sources and significantly more disturbed sources when compared to the nonmergers. Nonmergers have comparable proportions of C sources, CE sources, and jetted sources and contain all the E sources and significantly fewer disturbed sources as compared to the mergers. We believe that this is a result of the large-scale bulk motions of the ICM that are present in the merging environment (see discussion in \S\ref{sec:disc}).

\subsection{Comparing Radio and Optical AGN}
\label{sec:results:optragn}
In Table \ref{tab:agnclassdist}, we report the statistics of radio AGNs that are (a) also identified as an optical AGN (AGN$_{\mathrm{Rad-Opt}}$) and (b) only identified in the radio (AGN$_{\mathrm{Rad}}$) across both dynamical states. For the AGN$_{\mathrm{Rad-Opt}}$, we cross-reference our radio AGN sources with the \cite{Bilton2020} optical AGN catalog, and the sources that are in both samples are counted as AGN$_{\mathrm{Rad-Opt}}$. Conversely, the radio AGN in our sample that are not identified by \cite{Bilton2020} as an optical AGN and are classified as radio AGN only and are thus counted as  AGN$_{\mathrm{Rad}}$.

In comparing the proportions of AGN$_{\mathrm{Rad-Opt}}$ and AGN$_{\mathrm{Rad}}$ in mergers and nonmergers, we only find one marginally significant result: there are more AGN$_{\mathrm{Rad}}$ than AGN$_{\mathrm{Rad-Opt}}$ in merging systems taking into account the errors at a 2$\sigma$ level. In other words, the posterior probability that Pr(AGN$_{\mathrm{Rad}}$ $>$AGN$_{\mathrm{Rad-Opt}}$) within mergers is 0.96. Less significant but noteworthy, we find that mergers exhibit a marginally greater proportion of AGN$_{\mathrm{Rad}}$ at the 1.5$\sigma$ level as compared to nonmergers, or the posterior probability that Pr(M$>$NM) is 0.86. Next, comparison of AGN$_{\mathrm{Rad}}$ and AGN$_{\mathrm{Rad-Opt}}$ proportions within nonmergers yields a Pr(AGN$_{\mathrm{Rad}}$$>$AGN$_{\mathrm{Rad-Opt}}$) equivalent statistical significance less than one, where the Pr(AGN$_{\mathrm{Rad}}$ $>$AGN$_{\mathrm{Rad-Opt}}$) is 0.59. Lastly, the comparison of AGN$_{\mathrm{Rad}}$ and AGN$_{\mathrm{Rad-Opt}}$ across mergers and nonmergers, yields a  Pr(M$>$NM)  equivalent statistical significance of less than 1, where Pr(M$>$NM)  is 0.14.

\begin{table}
\caption {Distribution of Optical and Radio AGN across Cluster Dynamical States.}
  \renewcommand{\arraystretch}{1}
  \begin{tabularx}{0.47\textwidth}{X*{2}{>{\centering\arraybackslash}X}}
    \toprule
    & AGN$_{\mathrm{Rad-Opt}}$ & AGN$_{\mathrm{Rad}}$
    \\
     \toprule
    Tot$_{NM}$ & 19 & 18 \\
    Tot$_{M}$ & 6 & 11 \\
    \midrule
    P$_{\mathrm{NM}}$ & $0.51^{+0.08}_{-0.08}$  &  $0.49^{+0.08}_{-0.08}$ \\
    P$_{\mathrm{M}}$ & $0.35^{+0.12}_{-0.11}$&  $0.65^{+0.11}_{-0.12}$ \\
    \midrule
    $\mathrm{Pr(M>NM)}$ & 0.14 & 0.86 \\
    \bottomrule
   \end{tabularx}
  \label{tab:agnclassdist}
  \tablecomments{Columns are radio AGN coinciding with an optical AGN determined in \cite{Bilton2020} (AGN$_{\mathrm{Rad-Opt}}$) and only radio AGN (AGN$_{\mathrm{Rad}}$). Rows include the total number and the proportion with beta-binomial upper and lower bound uncertainties using a Jeffreys Bayesian prior across both dynamical states. $\mathrm{Pr(M > NM)}$ corresponds to the posterior probability that a specific classification is greater in mergers compared to nonmergers, discussed in Section \ref{sec:meth:beta}.}
\end{table}

\section{Discussion} \label{sec:disc}
This paper examines the population proportion of radio AGN morphologies and radio AGN and optical AGN across merging and nonmerging clusters. Our findings suggest that mergers have a statistically significant (at a 3$\sigma$ level) increased prevalence of disturbed radio sources and that mergers and nonmergers have a relatively equal prevalence of compact and jetted sources. Notably, a merging cluster environment enables large-scale bulk motion in the ICM, increases the potential for more galaxy$-$galaxy interactions, and heightens the overall disorder in the system. A dynamic environment such as that found in merging systems creates more opportunities for radio jets to interact with their surroundings and become bent \citep{Morris2022,ODea2023} or ``disturbed."

It has already been suggested in the literature that merging cluster systems can affect radio AGN morphology. There are historically two types of bent-tail radio AGN: wide-angle tails (WATs; which have a larger opening angle between the two radio jets) and narrow-angle tails (NATs; which have a smaller opening angle between the two radio jets). WATs seem to be fundamentally different from NATS in a variety of ways. The main difference is that WATs are cluster-center objects near the bottom of the cluster potential, and thus WAT host galaxies must be nearly at rest \citep{Owen1976,Quintana1982,Eilek1984}. Therefore, bent WAT morphology cannot be explained by ram pressure owing to the motion of the host galaxy. Instead, the WAT bent morphology has been posited to be a result of merging cluster$-$cluster systems \citep{Roettiger1993,Pinkney1994,Burns1998,Sakelliou2000,Burns2002} where the bulk scale motions of the ICM are responsible for the ram pressure that bends the tails. And even though there is still debate over the origin of hybrid morphology radio sources (intrinsic or environmental), an asymmetric environment has been posited as one origin \citep{Gopal-Krishna2000,Kapinska2017} that could be created in a cluster$-$cluster merger \citep{Gawronski2006}. Therefore, it is not unexpected that a merging system could cause a change in morphology of jetted systems as we see in this work.

We note that we only see HTs in merging systems (see tables in Appendix \ref{sec:app:Consensus3vs4}). HTs are extreme versions of BT systems and are a product of stronger ram pressure from either a denser medium or a higher velocity of the source ($P_{\mathrm{ram}} = \rho v^2$; \citealt{Miley1972,Terni2017}). Because only the most disturbed systems (HTs) are found in merging systems, this may indicate that merging systems allow for more opportunities for extreme ram pressure. An interesting follow-up would be to investigate the placement of these HT systems with respect to the cluster center.

Our findings suggest that mergers have a marginally significant increased prevalence of radio AGN without an optical-AGN counterpart. If we assume that AGN$_{\mathrm{Rad}}$ can be roughly equated to an inefficiently accreting engine or low-excitation radio galaxies (LERGs) and the AGN$_{\mathrm{Rad-Opt}}$ can be roughly equated to an efficiently accreting engine or high-excitation radio galaxies (HERGs \cite{Hardcastle2009,Buttiglione2010,Best2012}), then we see that there are more LERGs than HERGs in merging clusters. Though we do not do a full HERG/LERG classification in this analysis (this was not the goal of this work, so potential biases exist such as the fact that this sample does not have an S/N requirement for the spectral lines needed to classify HERG/LERGs), we note that it would be interesting future work. In fact, the environment affecting the incidence of LERGs versus HERGs is not unprecedented. A few early works find that in the local Universe LERGs occupy more dense environments than HERGs \citep{Prestage1988,Allington-Smith1993,Zirbel1997,Best2004}, which was confirmed by \cite{Miraghaei2017}. Similarly, \cite{Ineson2015} also find a difference between the environments of HERGs and LERGs. On the other hand, \cite{Massaro2019,Massaro2020} do not find a difference in HERG/LERG environments in the local Universe and note that the difference in environments of HERGs and LERGs seen in other works could be due to selection bias. 

\subsection{Discussion of Potential Sources of Uncertainty} \label{sec:disc:error}
There are several sources of error that we discuss here. An inherent source of error exists in the limitations of our classifications. Each source presents a unique radio emission profile, and while we aimed to encompass the bulk trends with the classification schemes, some intricacies may have eluded our classification process. Our morphological classification system does not encompass every type of radio galaxy morphology and was not designed to. We acknowledge the presence of jellyfish galaxies in our sample. Notably, we find examples of WC sources exhibiting faint extended emission that do not align exactly with the CE as defined here in this work but are noteworthy. In fact, two WC sources in A2255 are found in the existing literature to be ram-pressure-stripped galaxies \citep{Ignesti2023}. Further, we cross-match our sample with the LoTSS jellyfish catalog from \cite{Roberts2021} and find that four of our radio AGN are identified as jellyfish galaxies. We find that the jellyfish galaxies are equally spread between the WC, SC, CE, and E categories. However, we note that the LoTSS jellyfish catalog \citep{Roberts2021} only includes data from LoTSS DR2 and that up to half of clusters in this work have LoTSS data that have been taken since DR2. Similarly, we identified sources displaying FR-type morphologies that do not precisely fit FRM or FR I/ II categories, potentially due to image artifacts, yet they exhibit distinctive and intriguing morphologies. To maintain the scope of this project, we limit our results to only the original Zooniverse classification results.

An additional source of error that arises inherently in our classification-based project is disagreement. First, there are only 71 sources with complete agreement from all classifiers (consensus level 5), and the remaining 90 sources do not have complete agreement. On the other end of the spectrum within our data set, there were 26 sources that did not have a majority agreement (consensus level 2). Involving multiple classifiers was a necessary step to obtain robust classifications, even though it introduces some margin for error. There is a growing movement to move toward having multiple classifications or ``tags" per source in the radio classification community in order to better describe the complex nature of the morphology of radio galaxies \citep{Rudnick2021}. For this exploratory work, we chose to have only one classification per object, but that leaves room for ``error" in that we may not capture the full, complex morphology of the source. 

Another source of error is that of the dynamical state classification of the host cluster. It is difficult to obtain a binary classification of the dynamical state of clusters into ``merging" and ``relaxed." Galaxy clusters are a direct result of hierarchical merging systems evolving over cosmic time, and the clusters we examine are snapshots in time, each characterized by varying degrees of merging activity. There is an ongoing debate surrounding the optimal approach for determining the dynamical state of a galaxy cluster. Our study is based on the \cite{Bilton2020} data set, which classifies clusters as merging or nonmerging based on a kinematic analysis. However, existing literature suggests that certain clusters that were classified as a `nonmerging' cluster by \cite{Bilton2020} may in fact be undergoing merging processes based on X-ray observations that trace ICM turbulence \citep{Nulsen2013,Blanton2011,Drabent2015,Heng2016, Wen&Han2013}. Another way to classify the dynamical state of a cluster is based on the presence of central radio halos and radio relics, which can serve as an effective indicator of ongoing merger events. Notably, at least two clusters (A1367 and A1656) in this sample, classified as nonmerging using the \cite{Bilton2020} method, are identified as mergers based on radio halo observations \citep{Kim1990,Gavazzi1978,Bonafede2022,Ge2019,Zhang2023}. 

With these facts in mind, the clusters that \cite{Bilton2020} identified as merging and nonmerging via kinematics might not have that classification if one were to use a different type of analysis. Though it is difficult to get uniform, multiwavelength data for a large sample of clusters, and depending on the science goals, a multiwavelength approach would be the most robust best way to determine the dynamical state of a cluster.

A last source of error is the choice to use a global RMS for the cutout contours. The choice of a global RMS per cluster inherently may lead to the loss of smaller-scale emission for individual sources where the RMS may be different from the global RMS. Given our original dataset from the \cite{Bilton2020} catalog, which comprises over 1000 sources, we were confronted with a trade-off between efficiency and sensitivity. Consequently, emission of lower surface brightness, such as that associated with ram pressure stripping, may not be fully captured. Existing work suggests at least two WC sources from A2255 are ram-pressure-stripped galaxies \citep{Ignesti2023}. However, in \S\ref{sec:cutouts} we investigate the effect of choosing global versus a local RMS and find that our choice of a global, cluster RMS does not affect our results.

\section{Summary} \label{sec:summary}
In this work, we explored the effect of galaxy cluster mergers on the radio morphology of radio AGN. Starting from a sample of 33 galaxy clusters identified as merging (8) or nonmerging (25) by \cite{Bilton2020}, we assembled a sample of 12 galaxy clusters (8 merging and 4 nonmerging) that have LoTSS 144 MHz data. Our method to determine whether the merging cluster environment has an effect on radio AGN morphology was to have 5 experts classify the sources with a radio detection using Zooniverse, according to whether a source is weak compact (WC), strong compact (SC), compact extended (CE), extended (E), FR I or FR II, FR-morphology (FRM), bent-tail (BT), or a head-tail (HT). We combined the radio morphologies classes into C (WC+SC), CE, E, jetted (FR I + FR II + FRM), and disturbed (BT + HT), ( see \S\ref{sec:results:combinedClassifications}), and we obtain the following results:
\begin{itemize}
    \item Mergers do not exhibit a greater proportion of compact, compact extended, and jetted sources as compared to nonmergers (see \S\ref{sec:results:combinedClassifications}).
    \item Nonmergers contain all the extended classified sources in this study.
    \item Mergers have a statistically significant (3$\sigma$) higher population proportion of Disturbed sources than nonmergers (see \S\ref{sec:results:combinedClassifications}).
    \item Within merging clusters there is a marginally statistically significant (2$\sigma$) higher proportion of AGN that are identified only as radio AGN compared to radio AGN that are also identified as optical AGN (see \S\ref{sec:results:optragn}).
\end{itemize}
Overall, we find that the merging environment seems to have an effect on the radio morphology of radio sources in that there are more disturbed (bent and head-tails) sources in mergers. Our findings suggest the merging environment provides more opportunity for the jets to become disturbed.

We also investigated how the AGN detection fraction changes with wavelength (see \S\ref{sec:results:optragn}), in particular the fraction of sources that are detected in the radio versus optical. We obtain the following results:\begin{itemize}
    \item Within merging clusters there is a marginally statistically significant (2$\sigma$) higher proportion of AGN that are identified only as radio AGN compared to radio AGN that are also identified as an optical AGN (see \S\ref{sec:results:optragn}).
    \item Mergers do not exhibit greater proportions of radio AGNs and radio AGNs with an optical AGN detection compared to nonmergers.
\end{itemize}

This study presents significant opportunities for future work. One future direction is to follow a similar analysis to \cite{Bilton2020} and examine the spatial distribution of the classified sources in the phase-space diagram. Of particular interest is the potential trends of radio classification at certain spatial cuts within the cluster. A similar future direction would be to investigate the phase-space position of the AGN$_{\mathrm{Rad}}$ and AGN$_{\mathrm{Rad-Opt}}$. An additional future direction is to extend this work to a larger sample of clusters and have a spectrum of cluster dynamical states. A last future direction would be the further investigation of the "undetermined" or "nonconsensus" morphological classifications, as there is a significant variation in the morphologies in these morphology classes.

\section*{Acknowledgements}
We would like to thank the anonymous referee for their useful comments that made this work better. We would also like to thank Drs. Abhijeet Borkar and Paul Martini for helpful discussions. This work was made possible by the support of the National Radio Astronomy Observatory and Green Bank Observatory, which are facilities of the US National Science Foundation operated under cooperative agreement by Associated Universities, Inc. Y.A.G. is supported by US National Science Foundation (NSF) grant AST 22-06053. This publication uses data generated via the Zooniverse.org platform, development of which is funded by generous support, including a Global Impact Award from Google, and by a grant from the Alfred P. Sloan Foundation.

\vspace{5mm}
\facility{LOFAR (LoTSS)}

\software{astropy \citep{Astropy2018}}

\appendix
\label{Appendix}

\begin{table*}
\caption {Radio AGN Morphology Classifications with a Consensus Level of 3 and Above.}
  \centering
  \renewcommand{\arraystretch}{1.0} % Adjust row spacing
  \begin{tabularx}{\textwidth}{X*{8}{>{\centering\arraybackslash}X}}
    \toprule
    & WC & SC & CE & E & FRI+II & FRM & BT & HT  \\
    \toprule
    Tot$_{\mathrm{NM}}$ & 14 &  8 &  6 & 5 &      1 &   2 &  1 &  0 \\ % checked
    Tot$_{\mathrm{M}}$ & 6 &  3 &  1 & 0 &      2 &   0 &  1 &  4  \\ % checked
    \bottomrule
   \end{tabularx}
  \label{tab:345_stats}
    \tablecomments{Column names are shorthand for the classifications detailed in \S\ref{sec:classifications} weak compact (WC), strong compact (SC), compact extended (CE), extended (E), Fanaroff and Riley I+II (FRI+II), Fanaroff and Riley Morphology (FRM), bent-tail (BT), head-tail (HT), and uncertain (Unc). Rows detail the total, average (Avg), and normalized average (Norm) across mergers (M) and nonmergers (NM) for each morphology classification of consensus level of 3 after radio AGN delineation in section \ref{sec:agnVsSF}. }
\end{table*}

\begin{table*}
\caption {Radio AGN Morphology Classifications With a consensus Level of 4 and Above.} \label{tab:45_stats}
  \centering
  \renewcommand{\arraystretch}{1.0} % Adjust row spacing
  \begin{tabularx}{\textwidth}{X*{8}{>{\centering\arraybackslash}X}}
    \toprule
    & WC & SC & CE & E & FRI+II & FRM & BT & HT  \\
    \toprule
    Tot$_{\mathrm{NM}}$ & 10 &  6 &  5 & 2 &      1 &   1 &  0 &  0 \\ % checked
    Tot$_{\mathrm{M}}$ & 6 &  2 &  0 & 0 &      1 &   0 &  0 &  2 \\ % checked
    \bottomrule
   \end{tabularx}
    \tablecomments{Column names are shorthand for the classifications detailed in Section \S\ref{sec:classifications} weak compact (WC), strong compact (SC), compact extended (CE), extended (E), Fanaroff and Riley I+II (FRI+II), Fanaroff and Riley Morphology (FRM), bent-tail (BT), head-tail (HT). Rows detail the total across mergers (M) and nonmergers (NM) for each morphology classification of consensus level of 4 after radio AGN delineation. }
\end{table*}

\begin{deluxetable*}{lllrrlll}
\centering
 \tablecaption{Consensus Level 3 Sources}
\label{tab:example_data}
\tablehead{ \colhead{Name} & \colhead{Cluster} & \colhead{D$_{\mathrm{state}}$}& \colhead{RA}& \colhead{DEC}& \colhead{Radio$_{\mathrm{class}}$}& \colhead{RAGN}& \colhead{OAGN}}
\startdata
SDSS J031447.7+421321 & A0426 &      M &  48.698868 & 42.222702 &          WC &             False & False \\
SDSS J114041.6+202034 & A1367 &     NM & 175.173660 & 20.342951 &          WC &             False & False \\
SDSS J114224.4+200709 & A1367 &     NM & 175.602066 & 20.119303 &          CE &             False & False \\
SDSS J114358.1+204822 & A1367 &     NM & 175.992096 & 20.806389 &          WC &             False & False \\
SDSS J114358.9+200437 & A1367 &     NM & 175.995667 & 20.077028 &          CE &             False & False \\
SDSS J114527.6+204825 & A1367 &     NM & 176.365280 & 20.807171 &          CE &             False & False \\
SDSS J114548.8+203743 & A1367 &     NM & 176.453491 & 20.628616 &          WC &             False & False \\
SDSS J114612.1+202329 & A1367 &     NM & 176.550766 & 20.391647 &          SC &              True &  True \\
SDSS J115148.4+202727 & A1367 &     NM & 177.951752 & 20.457710 &          WC &             False & False \\
SDSS J125217.7+270507 & A1656 &     NM & 193.073837 & 27.085546 &          WC &             False &  True \\
SDSS J125351.5+285845 & A1656 &     NM & 193.464584 & 28.979395 &          WC &             False & False \\
SDSS J125359.1+262638 & A1656 &     NM & 193.496307 & 26.444042 &          WC &             False & False \\
SDSS J125455.1+272445 & A1656 &     NM & 193.729874 & 27.412704 &          WC &             False & False \\
SDSS J125547.8+281521 & A1656 &     NM & 193.949280 & 28.256104 &          WC &             False & False \\
SDSS J125651.1+265356 & A1656 &     NM & 194.213211 & 26.898890 &          WC &             False &  True \\
SDSS J125732.8+273637 & A1656 &     NM & 194.386826 & 27.610346 &          WC &             False & False \\
SDSS J125835.1+273547 & A1656 &     NM & 194.646606 & 27.596390 &          SC &             False & False \\
SDSS J125855.9+275000 & A1656 &     NM & 194.733185 & 27.833393 &          CE &             False & False \\
SDSS J125939.1+285343 & A1656 &     NM & 194.912979 & 28.895521 &          WC &             False & False \\
SDSS J130125.2+291849 & A1656 &     NM & 195.355286 & 29.313740 &          CE &              True &  True \\
 \enddata
 \tablecomments{The first 20 rows of the full sample of consensus level 3 sources. The full table is provided with this article. Columns are the SDSS name of the source \cite{Bilton2020} (Name); the name of the cluster that the source is in (Cluster); the dynamical state of the host cluster, D$_{state}$, which is either nonmerging (NM) or merging (M) as defined in \cite{Bilton2020}; R.A., Decl, the radio classification assigned through the Zooniverse classification done in this work (Radio class, see \S\ref{sec:classifications}); whether or not the source is classified as a radio AGN (RAGN, see \S\ref{sec:agnVsSF}); and whether or not the source is classified as an optical AGN in \cite{Bilton2020} (OAGN).}
\end{deluxetable*}

\section{Consensus Level 3 vs. Consensus Level 4 Results}
\label{sec:app:Consensus3vs4}

In Table \ref{tab:345_stats}, we present the radio AGN statistics for consensus level 3 after radio AGN delineation but before the combined classifications. In Table \ref{tab:45_stats}, we present the consensus level 4 statistics after radio AGN delineation and before combining classes. We determine that the results do not change when comparing consensus level 3 and 4. For example, when we compare the results for both consensus level 4 and consensus level 3 C, CE, E, or jetted sources, we find there is no difference and that both yield that mergers do not exhibit statistically significantly greater proportions of either C, CE, E, or jetted sources compared to nonmergers (in other words, Pr(M$>$NM) corresponds to a $\sigma \leq$ 1). Similarly, when we compare the results for both consensus level 4 and consensus level 3 disturbed sources, we again find there is no difference in that both yield that mergers exhibit marginally significant greater proportions of disturbed sources in consensus level 4 (where Pr(M$>$NM) is 0.982 which is equivalent to $2.4\sigma$ for the consensus level 4 sources). We note that in the consensus level 4 sources there are no CE in mergers. This differs from consensus level 3 sources where we observe only one CE source. We attribute the reduction of CE sources to a significant sample size decrease and is likely not physical in origin. Sample size plays a significant role in consensus level 4, as it leads to more values consistent with null. While consensus level 4 has a significantly smaller sample size, it still yields similar results as consensus level 3. While a change in consensus level does change the size of our sample, our results do not change. Thus, we decided to use the consensus level sample. 

\section{Zooniverse Results}
\label{sec:app:raw_zooniverse}

We provide a catalog of the 157 classification results following the aggregation of votes from the Zooniverse project but prior to combining into bulk group classifications (see \S\ref{sec:consensus_level} for details) with this work and display the first 20 lines in Table \ref{tab:example_data}. This does not include differentiation between the source of radio emission from a radio AGN or star formation. With the information from Table \ref{tab:example_data}, the reader can reproduce Table \ref{tab:345_stats} and the main results from this work that are presented in Table \ref{tab:combinedMorphologyStats} by including only those that are identified as radio AGN (see the methods of \S\ref{sec:agnVsSF}). To do this, the reader would filter Table \ref{tab:example_data} for only those with `RAGN' == True.

\bibliography{radioAGN_inMerg_lowz,software}{}
\bibliographystyle{aasjournal}

\end{document}

%% file: authors.tex
%% 
%ORDER TBD
%%
\author[0009-0007-9943-1183]{Mary Rickel}
\altaffiliation{co-first author}
\altaffiliation{Mary Rickel was a summer student at Green Bank Observatory}
\affiliation{Green Bank Observatory, P.O. Box 2, Green Bank, WV 24944, USA}
\affiliation{Department of Astronomy, The Ohio State University, 140 West 18th Avenue, Columbus, OH 43210, USA}
\affiliation{Department of Physics, University of Notre Dame, Notre Dame, IN USA}
\author[0000-0001-9793-5416]{Emily Moravec}
\altaffiliation{co-first author}
\affiliation{Green Bank Observatory, P.O. Box 2, Green Bank, WV 24944, USA}
\author[0000-0003-1432-253X]{Yjan A. Gordon}
\affiliation{Department of Physics, University of Wisconsin-Madison,1150 University Ave., Madison, WI 53706, USA}
\author[0000-0003-4223-1117]{Martin J. Hardcastle}
\affiliation{Department of Physics, Astronomy and Mathematics, University of Hertfordshire, College Lane, Hatfield, Hertfordshire AL10 9AB, UK}
\author[0000-0002-8250-9083]{Jonathon C. S. Pierce}
\affiliation{Department of Physics, Astronomy and Mathematics, University of Hertfordshire, College Lane, Hatfield, Hertfordshire AL10 9AB, UK}
\author[0000-0002-4780-129X]{Lawrence E. Bilton}
\affiliation{Institute of Physics, Pontifical Catholic University of Valpara\'iso, Brasil 2950, Valpara\'iso, Chile}
\affiliation{Institute of Physics \& Astronomy, University of Valpara\'iso, Blanco 951, Valpara\'iso, Chile}
\affiliation{Centre of Excellence for Data Science, Artificial Intelligence \& Modeling, The University of Hull, Cottingham Road, Kingston-Upon-Hull, HU6 7RX, UK}
\author[0000-0002-0692-0911]{Ian D. Roberts}
\affiliation{Department of Physics \& Astronomy, University of Waterloo, Waterloo, ON N2L 3G1, Canada}
\affiliation{Waterloo Centre for Astrophysics, University of Waterloo, 200 University Ave. W, Waterloo, ON N2L 3G1, Canada}
\affiliation{Leiden Observatory, Leiden University, P.O. Box 9513, 2300 RA Leiden, The Netherlands}